
\input phyzzx
\date={August 1992}
\rightline {August, 1992}
\rightline {SUSX-TH-92/16.}
\title {Twisted Sector Yukawa Couplings For The $Z_3\times Z_3$
Orbifold}
\author{D. Bailin$^{a}$, \ A. Love$^{b}$ \ and \ W.A. Sabra$^{c}$\foot{Address
from January 1993, Department of Physics, Royal Holloway and Bedford New
College, University of London, Surrey, U.K.}}
\address {$^{a}$School of Mathematical and Physical
Sciences,\break
University of Sussex, \break Brighton U.K.}
\address {$^{b}$Department of Physics,\break
Royal Holloway and Bedford New College,\break
University of London,\break
Egham, Surrey, U.K.}
\address {$^{c}$Physics Department \break Birkbeck College,\break Malet
Street\break London WC1E 7HX}
\abstract{ The moduli dependent Yukawa couplings between twisted
sectors of the $Z_3\times Z_3$ orbifold are studied.}
\endpage
\REF\one{L.Dixon, J. A. Harvey, C. Vafa and E. Witten, Nucl. Phys.
B261 (1985) 678; B274 (1986) 285.}
\REF\two{ A. Font, L. E. Ibanez, F. Quevedo and A. Sierra, Nucl.
Phys. B331 (1991) 421.}
\REF\three{L. Dixon, D. Friedan, E. Martinec and S. Shenker, Nucl.
Phys. B282 (1987) 13.}
\REF\four{S. Hamidi, and C. Vafa, Nucl. Phys. B279 (1987) 465.}
\REF\five{ L. E. Ibanez, Phys. Lett. B181 (1986) 269.}
\REF\six{J.A. Casas and C. Munoz, Nucl. Phys. B332 (1990) 189}
\REF\seven {J.A. Casas, F. Gomez and C. Munoz,  Phys. Lett.
B251(1990) 99}
\REF\eight {J.A. Casas, F. Gomez and C. Munoz,  CERN preprint,
TH6194/91.}
\REF\nine{T. Kobayashi and N. Ohtsubu, Kanazawa
preprint, DPKU-9103.}  \REF\ten {T.T Burwick, R. K. Kaiser and H. F.
Muller, Nucl. Phys. B355 (1991) 689}
\REF\eleven{D. Bailin and A. Love, Phys. Lett.
B260 (1991) 56.}
\REF\twelve{D. J. Gross, J. A. Harvey, E. Martinec and R. Rohm,
Nucl. Phys. B267 (1986) 75.}
\REF\thirteen {D. Gepner,
Nucl. Phys. B296 (1988) 757.}
\REF\fourteen {A. Font, L. E. Ibanez, F. Quevedo, Phys. Lett.
B217 (1989) 34; 272; \hfill \break
A. Font, L. E. Ibanez, M. Mondragon, F. Quevedo
and G. G. Ross, Phys. Lett. \ B227 \ (1989) \ 34; \hfill \break
A. Font, L. E. Ibanez, F. Quevedo and A.Sierra, Nucl. Phys. B337 (1990)
119.}
\REF\fifteen {J. Distler and B. Greene, Nucl. Phys. B309 (1988)
295.}
\REF\sixteen {A. B. Zamolodchikov and V. A. Fateev, Sov. Phys. JETP
62 (1984) 215.}

A comparison of orbifold compactified string
theory models [\one,\two] with observation will require amongst
other things a knowledge of the Yukawa couplings. Particularly
important is the moduli dependence of Yukawa couplings, which can
arise amongst twisted sector states [\three, \four] because their
exponential dependence on orbifold \lq\lq moduli" may have a
bearing on hierarchies [\five]. Twisted sector Yukawa couplings have
been investigated for the $Z_3$ orbifold [\three-\six], for the
$Z_7$ orbifold [\seven] and, more recently, for all the $Z_N$
orbifolds [\eight-\ten].
It is our purpose here to study twisted sector Yukawa couplings for
the $Z_3\times Z_3$ orbifold, which has also been used [\two] in
the computation of potentialy realistic models.
In the first instance the $Z_3\times Z_3$ orbifold may be realised (see,
for example, Ref.[2]) by first constructing an $[SU(3)]^3$ torus
using the identifications under translations
$$X_i\sim X_i+e_i, \qquad i=1,2,3\eqn\tran$$
and
$$X_i\sim X_i+{\tilde e}_i, \qquad i=1,2,3\eqn\trans$$
with  $$e_i=1,\quad {\tilde e}_i=e^{2\pi i/3},\eqn\transf$$for all
values of $i$, where the 3-complex coordinates $X_i$ define the
six-dimensional compact manifold. The orbifold with point group
$Z_3\times Z_3$ is then obtained by identifying points on the torus
under the rotations
$$\theta : X_1\rightarrow e^{2\pi i/3}X_1, \quad X_2\rightarrow
X_2,\quad X_1\rightarrow e^{4\pi i/3}X_3\eqn\orb$$and
$$\omega : X_1\rightarrow X_1,\quad  X_2\rightarrow e^{2\pi
i/3}X_2,\quad \quad X_3\rightarrow e^{4\pi i/3}X_3.\eqn\orbi$$
The action of the point group on the lattice is then
$$\theta e_1={\tilde e}_1,\quad \theta e_2=e_2,\quad
\theta e_3=-e_3-{\tilde e}_3,\quad\theta{\tilde
e}_1=-e_1-{\tilde e}_1,\quad \theta{\tilde e}_2={\tilde
e}_2,\quad\theta{\tilde e}_3= e_3\eqn\point$$
$$\omega e_1=e_1,\quad\omega e_2={\tilde e}_2,\quad\omega
e_3=-e_3-{\tilde e}_3,\quad\omega{\tilde e}_1={\tilde e}_1,\quad
\omega{\tilde e}_2=-e_2-{\tilde e}_2,\quad\omega{\tilde
e}_3=e_3\eqn\points$$ More generally, the original (rigid) lattice
\transf\ may be deformed in ways which preserve the action \point\
and \points\ of the point group. To obtain the most general choice
of the lattice compatible with the point group we must require that
all the scalar products $e_i.e_j$, $e_i.{\tilde e}_j$ and
${\tilde e}_i.{\tilde e}_j$ are preserved by the transformations
\point\ and \points. If we write
$$e_i.e_j=\mid e_i\mid\mid e_j\mid \cos\theta_{ij}\eqn\sc$$
$$e_i.{\tilde e}_j=\mid e_i\mid\mid {\tilde e}_j\mid
\cos\theta_{i{\tilde j}}\eqn\sca$$ and
$${\tilde e}_i.{\tilde e}_j=\mid {\tilde e}_i\mid\mid {\tilde
e}_j\mid \cos\theta_{{\tilde i}{\tilde j}}\eqn\scal$$
we find that
$$\mid {\tilde e}_i\mid=\mid e_i\mid,\qquad i=1,2,3\eqn\scala$$ and
that all angles between distinct basis vectors are fixed to be
$\pi/2$ with the exception of $\theta_{1\tilde 1}$, $\theta_{2\tilde
2}$ and
 $\theta_{3\tilde 3}$ which are fixed to be $2\pi/3$
$$\cos \theta_{1\tilde 1}=\cos \theta_{2\tilde 2}=
 \cos\theta_{3\tilde 3}=-{1\over2}\eqn\scalar$$
Thus we may take the 3 independent deformations of the lattice
compatible with the point group to be
$$R_i\equiv \mid e_i\mid, \qquad i=1,2,3\eqn\scalars$$
In general, left chiral massless states occur in the $\theta$,
$\theta^2$, $\omega$, $\omega^2$, $\theta\omega^2$, $\theta^2\omega$
and $\theta\omega$ twisted sectors, and the point group selection
rules and $H$-momentum selection rules restrict [\two] the allowed
couplings amongst these sectors to the forms
$$DDD,\ \bar ABC,\ A\bar B\bar C,\ ACD,\ B\bar CD\ \hbox{and}\ \bar
A\bar BD\eqn\yukawa$$where we use $A$, $\bar A$, $B$, $\bar B$, $C$,
$\bar C$ and $D$, respectively, to denote the twisted sectors listed
above. The fixed tori and fixed points for the twisted sectors are
readily obtained using the action \point\ and \points\ of the
generators of the point group on the deformed lattice. The $\theta$
twisted sector $(A)$ states are associated with the 9
inequivalent fixed tori given by
$$f_{\theta}={m_1\over 3}(2e_1+\tilde e_1)+{m_3\over 3}(e_3-\tilde
e_3)+a_2e_2+b_2\tilde e_2,\quad m_1,m_2=0,\pm 1\eqn\net$$
where $a_2$ and $b_2$ are arbitrary. These are fixed tori of the
space group elements $\Big(\theta, l(\theta)\Big)$ where
$$l(\theta)=(I-\theta)f_\theta+(I-\theta)\Lambda=m_1e_1+m_3e_3+
(I-\theta)\Lambda,\eqn\fix$$
where $\Lambda$ signifies an arbitrary lattice vector. The
$\theta^2-$twisted sector $(\bar A)$ states are associated with the
same fixed tori $f_\theta$ but with the space group elements
$\Big(\theta^2, l(\theta^2)\Big)$ where
$$l(\theta^2)=(I-\theta^2)f_\theta+(I-\theta^2)\Lambda=-m_1e_1-m_3e_3+
(I-\theta^2)\Lambda.\eqn\fixe$$
Similarly, the $\omega$ and $\omega^2-$twisted sector states
$(B\ \hbox{and}\ \bar B)$ are associated with 9 inequivalent fixed
tori $$f_\omega={n_2\over3}(2e_2+\tilde e_2)+{n_3\over3}(e_3-\tilde
e_3)+c_1e_1+d_1\tilde e_1,\quad n_2, n_3=0,\pm 1,\eqn\fixed$$
where $c_1$ and $d_1$ are arbitrary. These are fixed tori of the
space group elements $\Big(\omega, l(w)\Big)$ with
$$l(\omega)=(I-\omega)f_\omega+(I-\omega)\Lambda=n_2e_2+n_3e_3+
(I-\omega)\Lambda\eqn\sp$$and $\Big(\omega^2, l(\omega^2)\Big)$ with
$$l(\omega^2)=(I-\omega^2)f_{\omega^2}+(I-\omega^2)\Lambda=
-n_2e_2-n_3e_3+(I-\omega^2)\Lambda.\eqn\spa$$
Also the $\theta\omega^2$ and $\theta^2\omega-$twisted sector
states $(C\ \hbox {and}\ \bar C)$ are associated with the 9
inequivalent fixed tori$$f_{\theta\omega^2}={p_1\over3}(2e_1+\tilde
e_1)+{p_2\over3}(e_2-\tilde e_2)+g_3e_3+h_3\tilde e_3,\quad p_1,
p_2=0,\pm 1\eqn\spac$$
where $g_3$ and $h_3$ are arbitrary. These are fixed tori of the
space group elements $\Big(\theta\omega^2, l(\theta\omega^2)\Big)$
with$$l(\theta\omega^2)=(I-\theta\omega^2)f_{\theta\omega^2}+
(I-\theta\omega^2)\Lambda=p_1e_1+p_2e_2+
(I-\theta\omega^2)\Lambda\eqn\space$$ and $\Big(\theta^2\omega,
l(\theta^2\omega)\Big)$ with
$$l(\theta^2\omega)=(I-\theta^2\omega)f_{\theta^2\omega}+
(I-\theta^2\omega)\Lambda=-p_1e_1-p_2e_2+
(I-\theta^2\omega)\Lambda\eqn\spaces$$
Finally, the $\theta\omega$ twisted sector states $(D)$ are
associated with the 27 inequivalent fixed points
$$f_{\theta\omega}={1\over3}\sum_{i=1}^3
r_i(2e_i+\tilde e_i),\qquad r_1,r_2,r_3=0,\pm
1\eqn\gr$$ with corresponding space group elements
$\Big(\theta\omega, l(\theta\omega)\Big)$ where
$$l(\theta\omega)=\sum_{i=1}^3
r_ie_i+(I-\theta\omega)\Lambda.\eqn\gro$$ All of these fixed tori and
fixed points reduce to those of [\eleven] in the case of the rigid
lattice of \transf. The space group selection rules for couplings
amongst these twisted sectors are identical in form to those given
in Ref. [11] for the rigid lattice case.

The values of the Yukawa couplings amongst the twisted sectors
just discussed are determined in detail by three-point functions
involving fermionic and bosonic string degrees of freedom.
However, the crucial dependence on the deformation parameters
(moduli) and the particular fixed points and fixed tori is
entirely contained in (bosonic) twist field correlation functions
[\three, \four] of the type
$$Z=\prod_{i=1}^3<\sigma_\alpha^i(z_1, \bar z_1)
\sigma_\beta^i(z_2, \bar z_2)\sigma_\gamma^i(z_3, \bar
z_3)>\eqn\cor$$ where $\alpha,$ $\beta$ and $\gamma$ label the
twisted sectors at the particular fixed points and fixed tori
involved, and the index $i$ distinguishes the twist fields
associated with the complex coordinates $X_i$, $i=1,2,3.$
In the case of the $DDD$ coupling, the discussion is identical up
to a point to that of the $Z_3$ orbifold [\four-\six] because
$\theta\omega$ is identical to the point group element generating
$Z_3$ and the same rigid lattice is involved. However, whereas the
deformation parameters (moduli) for the $Z_3$ orbifold contains
angles as well as moduli, the deformation paramters which enter the
Yukawa couplings for the $Z_3\times Z_3$ orbifold are restricted to
$R_1$, $R_2$ and $R_3$ of \scalars.

Let the three fixed points involved be labelled by $r_i^1,$
$r_i^2,$ and $r_i^3,$ $i=1,2,3,$ respectively, in the notation of
\gr. The space group selection rules [\three, \four, \eleven]
requires
$$\sum_{J=1}^3 r_i^J=0\ (\hbox{mod}\ 3),\quad i=1,2,3,\eqn\core$$
If we define
$$d_i=r_i^1-r_i^2\eqn\corel $$
then we can write the leading exponential in the $DDD$ Yukawa
couplings as
$$Z_{DDD}\sim \exp\Big(-{1\over 2\pi\sqrt
3}\sum_i\Delta_i^2R_i^2\Big)\eqn\corell$$where $$\Delta_i=d_i, \quad
\hbox{for}\ d_i=0,\pm 1\eqn\corella$$
and
$$\Delta_i=d_i\mp 3, \quad
\hbox{for}\ d_i=\pm 2.\eqn\corellat$$
For the $A\bar B\bar C$ Yukawa coupling we need to evaluate
$$Z_{A\bar B\bar C}=\prod_{i=1}^3<\sigma_\theta^i(z_1, \bar z_1)
\sigma_{\omega^2}^i(z_2, \bar z_2)\sigma_{\theta^2\omega}^i(z_3, \bar
z_3)>\eqn\corellati$$ where the index labelling the fixed torus for
each twist field has been suppressed. Because $\theta$ only rotates
$X_1$ and $X_3$, $\omega^2$ only rotates $X_2$ and $X_3$, and
$\theta^2\omega$ only rotates $X_1$ and $X_2$, some of the twist
fields are the identity and $Z_{A\bar B\bar C}$ reduces to
$$\eqalign{Z_{A\bar B\bar C}=&<\sigma_\theta^1(z_1, \bar z_1)
\sigma_{\theta^2\omega}^1(z_3, \bar
z_3)><\sigma_{\omega^2}^2(z_2, \bar
z_2)\sigma_{\theta^2\omega}^2(z_3, \bar z_3)>\cr
&\times <\sigma_\theta^3(z_1, \bar z_1)
\sigma_{\omega^2}^3(z_2, \bar z_2)>\cr}\eqn\corellatio$$
Thus, only 2-point functions for twist fields are involved (which
can be normalized to 1) and no dependence on moduli or the specific
fixed tori involved arises. A similar argument applies for the
$\bar ABC$ Yukawa coupling.

For the $ACD$ coupling the situation is more interesting. In that
case, we have to evaluate
$$Z_{ACD}=\prod_{i=1}^3<\sigma_\theta^i(z_1, \bar z_1)
\sigma_{\theta\omega^2}^i(z_2, \bar z_2)\sigma_{\theta\omega}^i(z_3,
\bar z_3)>.\eqn\corellation$$
In view of the fact that $\theta$ only rotates $X_1$ and $X_3$ and
$\theta\omega^2$ only rotates $X_1$ and $X_2$ this reduces to
$$\eqalign{Z_{ACD}=&<\sigma_\theta^1(z_1, \bar z_1)
\sigma_{\theta\omega^2}^1(z_2, \bar z_2)\sigma_{\theta\omega}^1(z_3,
\bar z_3)><\sigma_{\theta\omega^2}^2(z_2, \bar
z_2)\sigma_{\theta\omega}^2(z_3, \bar
z_3)>\cr &\times<\sigma_\theta^3(z_1, \bar
z_1)\sigma_{\theta\omega}^3(z_3, \bar z_3)>.\cr}\eqn\le$$
The last two factors can be normalized to 1, but the first factor
is non-trivial and can be calculated using the methods of Ref.
[\three]. The three twist field correlation function
$$Z_1=<\sigma_\theta^1(z_1, \bar z_1)
\sigma_{\theta\omega^2}^1(z_2, \bar z_2)\sigma_{\theta\omega}^1(z_3,
\bar z_3)>\eqn\leb$$
factors into a quantum piece $Z_{qu}$ and a classical piece with all
the dependence on the moduli and the particular fixed points and
fixed tori involved contained in the classical piece.
$$Z_1=Z_{qu} \sum_{X_{cl}}e^{-S_{cl}},\eqn\leba$$
where the classical action is
$$S_{cl}={1\over\pi}\int d^2z\Big
({\partial X_1\over \partial z}{\partial \bar X_1\over \partial
\bar z}+{\partial X_1\over \partial \bar z}{\partial \bar X_1\over
\partial z}\Big).\eqn\leban$$
Because of the string equations of motion
$${\partial^2 X_1\over \partial z\partial\bar z}=0,\eqn\lebano$$
${\partial X_1/\partial z}$ and ${\partial X_1/ \partial
\bar z}$ are functions of $z$ and $\bar z$ alone, respectively,
which have to be chosen to respect the boundary conditions at
$z_1$, $z_2$ and $z_3$ implicit in the operator product expansions
with the twist fields. Here, the relevant operator product
expansions are
$$\eqalign{{\partial X_1\over \partial z}\sigma_\theta^1(z_1,\bar
z_1)&\sim (z-z_1)^{-2/3},\cr
{\partial X_1\over \partial z}\sigma_{\theta\omega^2}^1(z_2,\bar
z_2)&\sim (z-z_2)^{-2/3},\cr
{\partial X_1\over \partial z}\sigma_{\theta\omega}^1(z_3,\bar
z_3)&\sim (z-z_3)^{-2/3},\cr}\eqn\lebanon$$
and
$$\eqalign{{\partial X_1\over \partial\bar z}\sigma_\theta^1(z_1,\bar
z_1)&\sim (\bar z-\bar z_1)^{-1/3},\cr
{\partial X_1\over \partial\bar
z}\sigma_{\theta\omega^2}^1(z_2,\bar z_2)&\sim (\bar z-\bar
z_2)^{-1/3},\cr
{\partial X_1\over \partial
\bar z}\sigma_{\theta\omega}^1(z_3,\bar z_3)&\sim (\bar z-\bar
z_3)^{-1/3}.\cr}\eqn\br$$
Correspondingly,
${\partial X_1/\partial z}$ and ${\partial X_1/\partial\bar
z}$ are of the form
$${\partial X_1\over \partial z}=
a_1(z-z_1)^{-2/3}(z-z_2)^{-2/3}(z-z_3)^{-2/3}\eqn\bri$$
and
$${\partial X_1\over \partial\bar z}=
b_1(\bar z-\bar z_1)^{-1/3}(\bar z-\bar
z_2)^{-1/3}(\bar z-\bar z_3)^{-1/3}.\eqn\brit$$
Only the holomorphic field ${\partial X_1/\partial z}$ is an acceptable
classical solution, because ${\partial X_1/\partial \bar z}$ gives a divergent
contribution to the classical action.  The contribution of ${\partial
X_1/\partial z}$ is of the form
$$S_{cl}=\Big[{\Gamma(1/3)\over \Gamma(2/3)}\Big]^3 \mid
a_1\mid^2\mid -z_3\mid^{-4/3}\eqn\briti$$
where we have used $SL(2,C)$ invariance to set
$$z_1=0, \quad z_2=1, \quad z_3=\infty\eqn\britia$$
and the integral has been evaluated with the aid of Appendix A of
ref. [12]. The allowed values of $a_1$ are determined by the global
monodromy condition [\three]
$$\oint_{\cal C}dz {\partial X_1\over\partial z}=v_1\eqn\britian$$
where $\cal C$ is a closed contour around which $X_1$ is shifted by
$v_1$ but not rotated. In the present case, the contour may be
chosen to be the contour of fig. 1, where the point $z_1=0$  is
encircled once anti-clockwise and the point $z_2=1$ is encircled
once clockwise. The relevant integral is an integral of the type
[\nine]
$$\oint_{\cal C}dz z^{-(1-k_1w)}(z-1)^{-(1-k_2w)}=-2i\sin(k_1k_2\pi
w){\Gamma(k_1w)\Gamma(k_2w)\over \Gamma(k_1w+k_2w)}\eqn\eng$$
where the contour $\cal C$ is such that it encircles
$z=0$ $k_2$ times anti-clockwise and $z=1$ $k_1$ times clockwise.
Thus,
$$a_1={i\over\sqrt3}{\Gamma(2/3)\over[\Gamma(1/3)]^2}(-z_3)
^{2/3}v_1\eqn\engl$$Moreover, the shift $v_1$ on $X_1$ for the
contour of fig. 1 is obtained as the component in the $X_1$-plane
(corresponding to $e_1$ and $\tilde e_1$ ) of the product of space
group elements
$\Big(\theta,(I-\theta)f_\theta+(I-\theta)\Lambda\Big)
\Big(\theta^2\omega,(I-\theta^2\omega)f_{\theta\omega^2}+
(I-\theta^2\omega)\Lambda'\Big)$
where $\theta^2\omega$ is $(\theta\omega^2)^{-1}$. Consequently
$v_1$ is of the form
$$v_1=(m_1-p_1)e_1+(I-\theta)\Lambda+(I-\theta^2\omega)\Lambda'
\eqn\engla$$
where the notations of \net\ and \spac\ for the fixed tori are being
used, and $\Lambda$ and $\Lambda'$ denote arbitrary lattice vectors.
Combining \leba, \leban, \briti, \engl\ and \engla\ and
recalling that $\mid e_1\mid=R_1$, we find the leading order
behaviour for the $ACD$ Yukawa coupling
$$\eqalign{Z_{ACD}&\sim 1,\qquad \hbox{for}\ m_1-p_1=0,\cr &\sim
\exp\Big(-{R_1^2\over 2\pi\sqrt 3}\Big),\qquad \hbox{for}\
m_1-p_1=\pm 1,\pm 2.\cr}\eqn\englan$$
A similar discussion applies to the $B\bar CD$ and $\bar A\bar BD$
Yukawa couplings. For the $B\bar CD$ coupling, the relevant
quantity is
$$Z_{B\bar CD}=<\sigma^2_\omega(z_1, \bar z_1)
\sigma^2_{\theta^2\omega}(z_2, \bar
z_2)\sigma^2_{\theta\omega}(z_3, \bar z_3)>\eqn\england$$
and the leading behaviour is
$$\eqalign{Z_{B\bar CD}&\sim 1,\qquad \hbox{for}\ n_2+p_2=0,\cr
&\sim \exp\Big(-{R_2^2\over 2\pi\sqrt 3}\Big),\qquad \hbox{for}\
n_2+p_2=\pm 1,\pm 2\cr}\eqn\scot$$
with the notation of \fixed\ and \spac\ for the fixed tori.
For the $\bar A\bar BD$ coupling the relevant quantity is
$$Z_{\bar A\bar BD}=<\sigma^3_{\theta^2}(z_1, \bar z_1)
\sigma^3_{\omega^2}(z_2, \bar
z_2)\sigma^3_{\theta\omega}(z_3, \bar z_3)>\eqn\scotl$$
and the leading behaviour is
$$\eqalign{Z_{\bar A\bar BD}&\sim 1,\qquad \hbox{for}\ n_3-m_3=0,\cr
&\sim \exp\Big(-{R_3^2\over 2\pi\sqrt 3}\Big),\qquad \hbox{for}\
n_3-m_3=\pm 1,\pm 2\cr}\eqn\scotla$$
with the notation of \net\ and \fixed\ for the fixed tori.
It is of some interest to compare the behavior of the $Z_3\times
Z_3$ orbifold Yukawa couplings just obtained with those of the
$1^9$ Gepner model [\thirteen] which is generally believed to
correspond to the $Z_3\times Z_3$ orbifold (without Wilson lines)
at an enhanced symmetry point where $R_1=R_2=R_3$. Elsewhere
[\eleven] we have already made identifications between the massless
states of these two models and checked that the space group
selection rules of the orbifold and the various selection rules of
the Gepner model lead to the same Yukawa couplings being non-zero.
Using these identifications and the connection [\fifteen, \sixteen]
between $N=2$ superconformal models and $SU(2)$ WZNW models, it is
not difficult to show that the couplings consistent with
the selection rules obtained from the $1^9$ Gepner model are of
the following form.
For the $DDD$ couplings, and with $d_i$ as in \corel, the non-zero
couplings are given by
$$b^9\qquad \hbox{for}\ d_1=\pm 1,\pm 2,\  d_2=\pm 1,\pm 2,\  d_3=\pm
1,\pm 2\  \hbox{etc}\eqn\scotlan$$
Other choices of the $d_i$ [\eleven] are not consistent with the
$U(1)^6$ selection rule which is present at the enhanced symmetry point.
For the $A\bar B\bar C$ couplings, all couplings are given by
$b^9$, and the non-zero $ACD$ couplings are given by
$$b^9\qquad \hbox{for}\ m_1-p_1=\pm 1, \pm 2\eqn\scotland$$

Other choices of the $m_1-p_1$ are not consistent with the $U(1)^6$
selection rule at the enhanced symmetry point.  In these expressions,
and in terms of the quantum numbers
$\pmatrix{\ &j&\ \cr m&\ &\bar m\cr}$ of $SU(2)$ WZNW primary fields,
we have used the following notation
$$b=<\pmatrix{\ &0&\ \cr 0&\ &0\cr}^2\pmatrix{\ &1\over2&\ \cr
1\over2&\ &1\over2&}>,\eqn\scotlands$$
(The couplings $\bar ABC$, $B\bar CD$ and $\bar A\bar BD$ are
similar)
It can be seen that the above expressions for the $DDD$, $A\bar
B\bar C$ and $ACD$ couplings are consistent with \corell\ at the
enhanced symmetry point $R_1=R_2=R_3$, with the independence of
the non-zero $A\bar B\bar C$ couplings from the particular fixed
tori involved, and with \englan.
\vskip 2cm
\centerline{\bf ACKNOWLEGEMENTS} We wish to thank J.A. Casas for a
helpful communication.  This work was supported in part by
S.E.R.C.
\vskip 1cm
\refout
\vskip 1cm
\centerline{\bf FIGURE CAPTION} \hfill \break Fig.1 \ The contour {\cal C}
used in the global monodromy condition.
\bye